\documentclass[english]{revtex4-2}
\usepackage[T1]{fontenc}
\usepackage[latin9]{inputenc}
\usepackage{geometry}
\usepackage{color}
\usepackage{amsmath}
\usepackage{amssymb}
\usepackage{setspace}

\makeatletter
\@ifundefined{date}{}{\date{}}

\usepackage{amsfonts}
\usepackage{babel}

\makeatother

\usepackage{babel}
\begin{document}
\title{Phase-retrieval from Bohm's equations}
\author{Carlos Alexandre Brasil,}
\affiliation{Departamento Acadêmico de Ciências da Natureza (DACIN), Campus Cornélio
Procópio (CP), Federal University of Technology -- Parana (UTFPR),
Avenida Alberto Carazzai 1640, 86300-000, Cornélio Procópio, PR, Brazil}
\affiliation{São Carlos Institute of Physics (IFSC), University of São Paulo (USP),
PO Box 369, 13560-970, São Carlos, SP, Brazil}
\email{carlosbrasil.physics@gmail.com}

\author{Miled Hassan Youssef Moussa and Reginaldo de Jesus Napolitano}
\affiliation{São Carlos Institute of Physics (IFSC), University of São Paulo (USP),
PO Box 369, 13560-970, São Carlos, SP, Brazil}
\begin{abstract}
We present an analytic method, based on the Bohmian equations for
quantum mechanics, for approaching the phase-retrieval problem in
the following formulation: By knowing the probability density $\left\vert \psi\left(\overrightarrow{r},t\right)\right\vert ^{2}$
and the energy potential $V\left(\overrightarrow{r},t\right)$ of
a system, how can one determine the complex state $\psi\left(\overrightarrow{r},t\right)$?
We illustrate our method with three classic examples involving Gaussian
states, suggesting applications to quantum state and Hamiltonian engineering. 
\end{abstract}
\maketitle

\section{\textit{Introduction} }

In this article we should address an issue which emerged with the
advent of the wave function concept: Is it possible to determine the
wave function $\psi\left(\vec{r},t\right)$, a complex entity, from
the experimentally measured density of probability $P_{r}\left(\vec{r},t\right)=\left\vert \psi\left(\vec{r},t\right)\right\vert ^{2}$,
that is a real entity by definition? This is an aspect of the general
mathematical/physical \emph{phase-retrieval problem} \citep{Fienup,Orlowski,Nakajima,Jaganathan,Liberman}
(or \emph{state determination problem} \citep{Aerts}), i.e., the
characterization of the phase of a general scalar field by its modulus
\citep{Gerchberg,Oppenheim}, not necessarily of quantum nature. Indeed,
we have early works of image signal/noise analysis on optical \citep{Fienup,Zernike1,Zernike2},
electronic microscopy and holography \citep{Gabor,Paganin} and, as
mentioned in \citep{Corbett1}, control theory and crystallography.
Already in the context of quantum wave function characterization,
we have both theoretical and experimental approaches on optical microscopy
\citep{Liberman}, general quantum optics \citep{Vaccaro,Nakajima},
cavity QED \citep{Bardroff} and ion traps \citep{Raymer}. A wide
general review can be found on Ref. \citep{Fienup,Oppenheim,Jaganathan,Paganin}.

The meaning of the solution of the wave equation proposed by Schrödinger
\citep{Ludwig,Schrodinger},

\begin{equation}
i\hbar\frac{\partial}{\partial t}\psi\left(\vec{r},t\right)=-\frac{\hbar^{2}}{2m}\nabla^{2}\psi\left(\vec{r},t\right)+V\left(\vec{r},t\right)\psi\left(\vec{r},t\right),\label{eqSchr3D}
\end{equation}
(where we follow the common definitions $\hbar=\frac{h}{2\pi}$, $h$
is the Planck's constant, $m$ the mass of the particle, $V\left(\vec{r},t\right)$
the energy potential over the particle, $\vec{r}$ is the position
vector and $t$ the time), was object of discussion by pioneers of
quantum mechanics like Schrodinger \citep{Ludwig,Schrodinger,Jammer}
himself, Dirac \citep{Dirac} and Born \citep{Ludwig,Born,Wheeler}
(the latter proposed the actually accepted probabilistic interpretation).
The work described here is motivated by the structure of the Bohm's
equations \citep{Bohm1,Bohm2}, that are totally equivalent to the
Schrödinger's one, and allow us to expose a general and analytic method
to solve the phase-retrieval problem.

The formulations of phase retrieval problem in quantum mechanics varied
according to time and authors, as already emphasized by Weigert \citep{Weigert}
- an early recapitulation on quantum mechanics was made on 40's by
Reichenbach \citep{Reichenbach}. Perhaps the first author \citep{Weigert,Moroz}
to approach this question was Pauli \citep{Pauli} who, after an analysis
of the one-dimensional case for a free particle, questioned himself
whether, in general, with the densities of probabilities $P_{r}\left(\vec{r},t\right)$
and $P_{p}\left(\overrightarrow{p},t\right)=\left\vert \widetilde{\psi}\left(\overrightarrow{p},t\right)\right\vert ^{2}$,
it would be possible to uniquely determine a function $\psi\left(\vec{r},t\right)$
that satisfies both and is \textquotedbl physically compatible\textquotedbl ,
i.e., it is as a solution of the Schrödinger equation (\ref{eqSchr3D})
for known $V\left(\vec{r},t\right)$. This formulation would become
known as \emph{the Pauli problem} \citep{Orlowski,Corbett1,Weigert,Gale,Pavicic,Richter,Corbett}.
Although Pauli posed the question for all time, several authors later
limited the issue to a particular instant $t_{0}$. Reichenbach \citep{Reichenbach}
talks about his own discussion with Bargmann about the ambiguity in
the definition of $\psi\left(\vec{r},t_{0}\right)$ if we know only
$P_{r}\left(\vec{r},t_{0}\right)$ and $P_{p}\left(\overrightarrow{p},t_{0}\right)$.
Bargmann showed we can have sets of functions on position $\left\{ \psi_{1}\left(\vec{r},t_{0}\right),\psi_{2}\left(\vec{r},t_{0}\right)\right\} $
and momentum $\left\{ \widetilde{\psi}_{1}\left(\overrightarrow{p},t_{0}\right),\widetilde{\psi}_{2}\left(\overrightarrow{p},t_{0}\right)\right\} $
representation with the same probability distribution for these observables,
i.e., $P_{r}\left(\vec{r},t_{0}\right)=\left\vert \psi_{1}\left(\vec{r},t_{0}\right)\right\vert ^{2}=\left\vert \psi_{2}\left(\vec{r},t_{0}\right)\right\vert ^{2}$
and $P_{p}\left(\overrightarrow{p},t_{0}\right)=\left\vert \widetilde{\psi}_{1}\left(\overrightarrow{p},t_{0}\right)\right\vert ^{2}=\left\vert \widetilde{\psi}_{2}\left(\overrightarrow{p},t_{0}\right)\right\vert ^{2}$,
but with $\left\vert \psi_{1}\left(\vec{u},t_{0}\right)\right\vert ^{2}\neq\left\vert \psi_{2}\left(\vec{u},t_{0}\right)\right\vert ^{2}$
and $\left\vert \widetilde{\psi}_{1}\left(\overrightarrow{u},t_{0}\right)\right\vert ^{2}\neq\left\vert \widetilde{\psi}_{2}\left(\overrightarrow{u},t_{0}\right)\right\vert ^{2}$
where $\hat{u}$ can be chosen in such a way that $\left[\hat{u},\hat{r}\right]\neq0$
and $\left[\hat{u},\hat{p}\right]\neq0$. Feenberg \citep{Reichenbach,Kemble}
presupposes knowledge in $t_{0}$ of the density of probability and
its time differentiation as sufficient condition to the definition
of $\psi\left(\vec{r},t_{0}\right)$. This hypothesis was later refuted
by Gale \textit{et al}. \citep{Gale}, who proposed that $P_{r}\left(\vec{r},t_{0}\right)$
and the probability current $\overrightarrow{j}\left(\vec{r},t_{0}\right)$
will be sufficient to determine $\psi\left(\vec{r},t_{0}\right)$
\emph{only} for a spinless particle. Corbett and Hurst \citep{Corbett1}
showed that, for $t_{0}$, bound states are uniquely determined by
$P_{r}\left(\vec{r},t_{0}\right)$ and $P_{p}\left(\overrightarrow{p},t_{0}\right)$,
but scattering states are not. Orlowski and Paul \citep{Orlowski}
proposed an iterative-approximative method to determine the state
in $t_{0}$ with the knowledge in this time of both densities of probabilities
- they applied the method for finite superpositions of photon-number
states. Richter \citep{Richter} modifies the Orlowski and Paul algorithm,
substituting the momentum distribution by the time differentiation
of $P_{r}\left(\vec{r},t_{0}\right)$. Followed by Vaccaro and Barnett
\citep{Vaccaro}, Bialynicka-Birula\citep{Bialynicka} proposed a
slight modification of Richter's algorithm, considering a finite superposition
of Fock states $\left\vert \psi\right\rangle =\underset{n}{\sum}A_{n}\exp\left(i\alpha_{n}\right)\left\vert n\right\rangle $
projected in the phase-space vector $\left\vert \phi\right\rangle \simeq\underset{n}{\sum}\exp\left(-in\phi\right)\left\vert n\right\rangle $,
resulting in $\psi\left(\phi\right)\simeq\underset{n}{\sum}A_{n}\exp\left(i\alpha_{n}\right)\exp\left(-in\phi\right)$:
they require the knowledge of $\left\vert \psi\left(\phi\right)\right\vert ^{2}$
and all the $A_{n}^{2}$.

Considering specifically a Stern-Gerlach apparatus \citep{Cohen},
Weigert \citep{Weigert} proposed that the wave function characterization
can be made with three measurements for the density of probability:
one along the $z$ direction, another on an infinitesimally rotated
one, $z^{\prime}$, and, at last, a measurement of the spin component
in a perpendicular orientation to the plane formed by $z$ and $z^{\prime}$.
More recently, Lundeen and collaborators \citep{Lundeen} proposed
a method for photon wave functions involving two measurements of two
complementary variables, the first measurement being a weak one \citep{Aharonov,Duck}.
They did not consider the time explicitly and, to Weigert \citep{Weigert,Weigert1},
the original Pauli problem was not solved yet. Other authors, instead
of considering a single particular fixed time, presuppose measures
referring to two or more instants. Kreinovitch \citep{Kreinovich}
and Wiesbrock \citep{Wiesbrock} proposed that position measurements
carried out at different times (within a known $V\left(\vec{r},t\right)$)
are sufficient to reproduce the \emph{initial} state of the ensemble
in a unique form. By its turn, Johansen \citep{Johansen} supposed
explicitly the knowledge of the potential $V\left(\vec{r},t\right)$
with the measurement of $P_{r}\left(\vec{r},t\right)$ for a discrete
number of time values spaced by a short time interval. As we can see,
the list of works in this field is very extensive \citep{Nakajima,Liberman,Aerts,Raymer,Corbett,Botero,Buzek,Freyberger,Paul}.

Our aim in this paper is to expose a general and analytic method,
based on Bohm's equations \citep{Bohm1,Bohm2}, to solve a phase-retrieval
problem. Here we determine the state phase as a function of both position
\emph{and} timeand our method requires \emph{only }the probability
amplitude $\left\vert \psi\left(\vec{r},t\right)\right\vert ^{2}$
\emph{and} $V\left(\vec{r},t\right)$, in a formulation similar to
that of Bardroff \textit{et al}. \citep{Bardroff} and Leonhardt and
Raymer \citep{Leonhardt}. We will not concern ourselves here with
how these quantities are measured. In the case of the probability
amplitude, for example, we can experimentally measure enough data
so that we can construct a grid in space and time with values of $P_{r}\left(\vec{r},t\right)$.
Then, using the numerical techniques involving smooth interpolation,
like cubic splines for 3D grids \citep{Sadikin}, we can calculate
the derivatives that are required in the equations we solve in our
procedure. After using the data to generate the smooth interpolated
functions, the whole prescrition here described can be applied without
any modication. Anyway, this issue is addressed in several of the
cited authors \citep{Orlowski,Vaccaro,Bardroff,Raymer,Reichenbach,Lundeen,Paul,Band1,Band2,Dormer}
and here we will simply assume that both quantities are known as functions
of both position and time. 

The Bohm's causal formulation of quantum mechanics, that emphasizes
the concept of quantum trajectories and had as precedents the ideas
of de Broglie \citep{Broglie} and Madelung \citep{Madelung}, has
a history of polemics, mainly related to the hidden variables question
\citep{Jammer,Jammer2}. However such a philosophical (or ideological)
controversy does not compromise the mathematical rigor of Bohm's equations
and our use of them will be merely mathematical and instrumental.
As we will see below, Bohm's equations emerge easily from Schrödinger's
one (\ref{eqSchr3D}), leading to two equations widely known before
the advent of quantum theory: the continuity and the Hamilton-Jacobi
ones \citep{FetterWalecka}, the latter being added by what is known
as the non-local quantum potential \citep{Holland,Wyatt}. The continuity
equation is already used to support some approaches \citep{Aerts,Paganin}
but, to our knowledge, before the present paper, only Ref. \citep{Coffey}
used the Hamilton-Jacobi equation for this purpose. However, in their
work the emphasis is on the numerical treatment of the experimental
data to estimate the probability density function, and they just suggest
the numerical solution of the Bohm's equations, with no details, to
obtain the phase. Here, on the other side, by the first time, we explain
in detail how to approach the equations, with purely analytic examples.

We illustrate our method with three unidimensional pure state cases,
described by Gaussians. In the first, we have an illustrative application
of our method to the free particle case and it can be compared with
the results on Ref. \citep{Aerts}. In the second case, we suppose
a general oscillating Gaussian to illustrate possible extensions of
our method for quantum state engineering \citep{Orlowski,Paul,Joshi,Vogel}.
Finally, in the third, we suppose an unusual probability density to
illustrate the application of our method to filter valid solutions
for the Schrödinger's equation for a given potential and to construct
specific Hamiltonians with desired solutions \citep{Pavicic,Pavicic1,Vogel,Almeida}.

The states considered here have characteristics that enable a wide
range of applications, since we can find numerous proposals have been
devoted to the phase retrieval of pure states \citep{Orlowski,Liberman,Vaccaro,Bardroff,Raymer,Weigert,Gale,Richter,Bialynicka,Lundeen,Weigert1,Wiesbrock,Freyberger,Paul},
considering unidimentional problems \citep{Orlowski,Weigert,Corbett,Lundeen,Wiesbrock,Johansen},
harmonic potential \citep{Orlowski,Richter,Bialynicka,Paul}, Gaussian
states \citep{Pavicic,Pavicic1} and time-dependent potentials \citep{Almeida}.

The article is organized as follows: Sec. II provides a mathematical
description of the method; Sec. III shows the examples; conclusions
and further sugestions for applications are presented in Sec. IV.

\section{\textit{The Method} }

The Bohm's equations \citep{Bohm1,Bohm2} result from the solution
of Eq. (\ref{eqSchr3D}) in the polar form $\psi\left(\vec{r},t\right)=R\left(\vec{r},t\right)\exp\left[\frac{i}{\hbar}S\left(\vec{r},t\right)\right]$,
where $P\left(\vec{r},t\right)=R^{2}\left(\vec{r},t\right)$ (omitting
the subscript $r$ on $P_{r}\left(\vec{r},t\right)$ since the momentum
distributions will not be used). If we substitute this polar form
on (\ref{eqSchr3D}), and separate the real and imaginary parts, we
obtain 
\begin{eqnarray}
\frac{\partial}{\partial t}S+\frac{\left(\vec{\nabla}S\right)^{2}}{2m}+V+U & = & 0,\label{eqBhom1}\\
\frac{\partial}{\partial t}P+\vec{\nabla}\cdot\left(P\frac{\vec{\nabla}S}{m}\right) & = & 0.\label{eqBohm2}
\end{eqnarray}
While Eq. (\ref{eqBohm2}) is the continuity equation, Eq. (\ref{eqBhom1})
is the Hamilton-Jacobi one \citep{FetterWalecka}, from which we can
define the canonical momentum $\overrightarrow{p}\left(\vec{r},t\right)=\overrightarrow{\nabla}S\left(\vec{r},t\right)$
and the non-local quantum potential 
\begin{equation}
U\equiv-\frac{\hbar^{2}}{4m}\left[\frac{\nabla^{2}P}{P}-\frac{\left(\vec{\nabla}P\right)^{2}}{2P^{2}}\right],\label{defU}
\end{equation}
which together with $V\left(\vec{r},t\right)$ gives the Bohm's total
potential $V_{B}\left(\vec{r},t\right)\equiv V\left(\vec{r},t\right)+U\left(\vec{r},t\right)$.
An important fact of Eq. (\ref{eqBhom1}) is that the density of probability
$P\left(\vec{r},t\right)$ is necessary and univocally related to
the phase $S\left(\vec{r},t\right)$. Then, if we know $V\left(\vec{r},t\right)$
and $P\left(\vec{r},t\right)$, we can determine $S\left(\vec{r},t\right)$
and, then, the complete state $\psi\left(\vec{r},t\right)$. Our method
involves the following steps:

$i)$ We first solve the continuity equation for $\overrightarrow{p}\left(\vec{r},t\right)$
by expanding the del operator, thus leading to a differential ordinary
equation with only the spatial differentiation on $\overrightarrow{p}\left(\vec{r},t\right)$:

\begin{equation}
\vec{\nabla}\cdot\vec{p}+\frac{\vec{\nabla}P}{P}\cdot\vec{p}=-\frac{m}{P}\frac{\partial}{\partial t}P\text{.}\label{ContEq}
\end{equation}
Solving it, we will obtain an expression for $\overrightarrow{p}\left(\vec{r},t\right)$
with an undetermined boundary condition $\Theta\left(t\right)$, which
is a function of time only.

$ii)$ Irrespectivelly to the first step, we then determine the Bohm's
potential $V_{B}\left(\vec{r},t\right)$.

$iii)$ We then solve the Hamilton-Jacobi equation (\ref{eqBhom1})
by differentiating it with respect to the spatial coordinates, to
obtain 
\begin{equation}
\frac{\partial}{\partial t}\vec{p}+\frac{1}{m}\left(\vec{p}\cdot\vec{\nabla}\right)\vec{p}+\vec{\nabla}V_{B}=0.\label{HJdiffs}
\end{equation}
By substituting $\overrightarrow{p}\left(\vec{r},t\right)$ found
previously from the first step into Eq. (\ref{HJdiffs}), we can determine
$\Theta\left(t\right)$, but now with another time boundary condition
$f\left(t\right)$ for $S\left(\vec{r},t\right)$.

$iv)$ Finally, we substitute the function $S\left(\vec{r},t\right)$
obtained from the latter step on the Hamilton-Jacobi equation (\ref{eqBhom1})
to determine $f\left(t\right)$, thus solving our problem apart from
an irrelevant constant factor.

\section{\textit{Illustrative Examples} }

We illustrate our method considering, as anticipated above, two unidimentional
examples: the free particle and the Harmonic oscillator.

\subsection{Free particle}

For a free particle on a Gaussian state we have the probability density

\begin{equation}
P\left(x,t\right)=\frac{1}{\sqrt{2\pi D(t)}}\exp\left[-\left(\frac{x-\frac{\left\langle p\right\rangle }{m}t}{\sqrt{2D(t)}}\right)^{2}\right]\text{.}\label{Pfreext}
\end{equation}
where we have defined the diffusion term 
\[
D(t)=\frac{\hbar^{2}}{4\left(\Delta p\right)^{2}}+\frac{\left(\Delta p\right)^{2}}{m^{2}}t^{2},
\]
with both the mean value $\left\langle p\right\rangle $ and the variance
$\Delta p$ of the mechanical momentum being constants. As expected
from our knowledge on the dynamics of a free quantum particle, the
diffusion term depends on the variance of the mechanical momentum.
Following the method exposed above, we start with the continuity equation
for $p\left(x,t\right)$, Eq. (\ref{ContEq}), whose solution is given
by 
\begin{equation}
p\left(x,t\right)=\frac{1}{D(t)}\left(\frac{\hbar^{2}\left\langle p\right\rangle }{4\left(\Delta p\right)^{2}}+\frac{\left(\Delta p\right)^{2}}{m}xt\right)+\Theta\left(t\right)\left\langle p\right\rangle \exp\left(\frac{x-2\frac{\left\langle p\right\rangle }{m}t}{2D(t)}x\right),\label{aux1}
\end{equation}
where the boundary condition $\Theta\left(t\right)$ remains to be
determined. The Bohm's potential for the free-particle becomes

\begin{equation}
V_{B}\left(x,t\right)=\frac{\hbar^{2}}{4mD(t)}\left[1-\frac{\left(x-\frac{\left\langle p\right\rangle }{m}t\right)^{2}}{2D(t)}\right],\label{aux2}
\end{equation}
and by substituting Eqs. (\ref{Pfreext}), (\ref{aux1}) and (\ref{aux2})
in the spatial derivative of the Hamilton-Jacobi equation (\ref{HJdiffs}),
we are left with

\begin{align}
\frac{d\Theta}{dt} & =\left(\frac{1}{D(t)}\frac{\hbar^{2}\left\langle p\right\rangle ^{3}}{4\left(\Delta p\right)^{2}}-\left(\Delta p\right)^{2}\right)\frac{\Theta\left(t\right)t}{m^{2}D(t)}\nonumber \\
 & -\frac{\left\langle p\right\rangle }{m}\left(x-\frac{\left\langle p\right\rangle }{m}t\right)D(t)\Theta^{2}\left(t\right)\exp\left[-\frac{x}{2D(t)}\left(x-2\frac{\left\langle p\right\rangle }{m}t\right)\right].\label{aux3}
\end{align}

Since, by hypothesis, $\Theta$ must be a function of time only, the
spatial differentiation of Eq. (\ref{aux3}) must be null and the
only admissible condition is $\Theta\left(t\right)\equiv0$. Consequently,
from the canonical momentum definition, it follows the action 
\begin{equation}
S\left(x,t\right)=\frac{1}{2D(t)}\left(\frac{\hbar^{2}\left\langle p\right\rangle }{2\left(\Delta p\right)^{2}}x+\frac{\left(\Delta p\right)^{2}}{m}x^{2}t\right)+f\left(t\right).\label{aux5}
\end{equation}
In order to determine $f\left(t\right)$ we substitute Eqs. (\ref{aux5})
and (\ref{aux2}) in the original Hamilton-Jacobi equation to find
\begin{equation}
f\left(t\right)=f\left(0\right)-\frac{\hbar^{2}\left\langle p\right\rangle ^{2}}{8m\left(\Delta p\right)^{2}}\frac{t}{D(t)}-\frac{\hbar}{2}\arctan\left[\frac{2\left(\Delta p\right)^{2}}{\hbar m}t\right],
\end{equation}
apart from an irrelevant overall factor $f\left(0\right)$.

\subsection{Harmonic oscillator}

We now turn to the harmonic oscillator case whose potential is given
by $V\left(x\right)=\left(m\omega^{2}/2\right)x^{2}$, where $\omega$
is the natural oscillation frequency. Here, instead of assuming a
completely definite probability density as we did in the case of a
free particle, we assume a general Gaussian form 
\begin{equation}
P\left(x,t\right)=\frac{\left\vert a\right\vert }{\sqrt{\pi}}e^{-a^{2}\left[x+b\cos\left(\omega t\right)\right]^{2}},\label{Pgauss}
\end{equation}
whose shape remains constant throughout its time evolution. The parameters
$a$ and $b$ are thus left to be determined under this coherence
assumption. Starting with the continuity equation (\ref{ContEq})
for $p\left(x,t\right)$, we thus derive the solution 
\begin{equation}
p\left(x,t\right)=\frac{m\omega}{a}\left(ab\sin\left(\omega t\right)+\Theta\left(t\right)e^{a^{2}\left[x^{2}+2bx\cos\left(\omega t\right)\right]}\right),\label{aux7}
\end{equation}
again with the boundary condition $\Theta\left(t\right)$. Going to
the Bohm's potential, from Eqs. (\ref{defU}) and (\ref{Pgauss})
we obtain $V_{B}\left(x,t\right)=U\left(x,t\right)+\frac{1}{2}m\omega^{2}x^{2}$,
where 
\begin{equation}
U\left(x,t\right)=\frac{\left(\hbar a\right)^{2}}{2m}\left\{ 1-a^{2}\left[x+b\cos\left(\omega t\right)\right]^{2}\right\} .\label{Uharm}
\end{equation}

By substituting Eqs. (\ref{Pgauss}) and (\ref{aux7}) on the expression
resulting from the spatial differentiation of the Hamilton-Jacobi
equation, we obtain

\begin{align}
\frac{d\Theta}{dt} & =\omega a\left[\left(\frac{\hbar a}{m\omega}\right)^{2}a^{2}-1\right]\left[x+b\cos\left(\omega t\right)\right]e^{-a^{2}\left[x^{2}+2bx\cos\left(\omega t\right)\right]}\nonumber \\
 & -\omega a\Theta\left(t\right)\left\{ ab^{2}\sin\left(2\omega t\right)+2\Theta\left(t\right)\left[x+b\cos\left(\omega t\right)\right]e^{a^{2}\left[x^{2}+2bx\cos\left(\omega t\right)\right]}\right\} \label{aux9}
\end{align}
Once again, by hypothesis $\Theta$ must be only a function of $t,$
such that the spatial derivative of Eq. (\ref{aux9}) becomes null
only if $\Theta\left(t\right)\equiv0$, and Eq. (\ref{aux7}) turns
out to be 
\begin{equation}
p\left(x,t\right)=m\omega b\sin\left(\omega t\right).\label{aux10}
\end{equation}
From the definition of the canonical momentum, it follows that 
\begin{equation}
S\left(x,t\right)=m\omega b\sin\left(\omega t\right)x+f\left(t\right),\label{aux11}
\end{equation}
where $a,$ $b$, and $f\left(t\right)$ remain to be determined.
By substituting Eqs. (\ref{aux10}) and (\ref{aux11}) on the Hamilton-Jacobi
equation, we obtain

\begin{equation}
\frac{df}{dt}+\frac{\left(\hbar a\right)^{2}}{2m}+\frac{b^{2}}{2}\left[m\omega^{2}\sin^{2}\left(\omega t\right)-\frac{\left(\hbar a\right)^{2}}{m}a^{2}\cos^{2}\left(\omega t\right)\right]+\left(m\omega^{2}-\frac{\left(\hbar a\right)^{2}}{m}a^{2}\right)\left[b\cos\left(\omega t\right)x+\frac{1}{2}x^{2}\right]=0\label{aux12}
\end{equation}
However, as $f$ must depend only of $t$, the last term on the left
hand side of Eq. (\ref{aux12}) must be null, what follows from the
complete determination of the parameter $a=\pm\sqrt{m\omega/\hbar}$,
leaving us with the simplified differential equation 
\begin{equation}
\frac{df}{dt}=\frac{1}{2}\hbar\omega\left(ab\right)^{2}\cos\left(2\omega t\right)-\frac{1}{2}\hbar\omega,
\end{equation}
whose solution is

\[
f\left(t\right)=f\left(0\right)+\frac{1}{4}\hbar\left(ab\right)^{2}\sin\left(2\omega t\right)-\frac{1}{2}\hbar\omega t\text{.}
\]

Whereas we have no constraint for the parameter $b$, which may assume
any real value, the initial value $f\left(0\right)$ makes no difference
on the final result. Then, we have two solutions for $\psi\left(x,t\right)$,
one coming from $a=\sqrt{m\omega/\hbar}$:

\begin{equation}
\begin{cases}
R\left(x,t\right) & =\left(\frac{m\omega}{\pi\hbar}\right)^{\frac{1}{4}}e^{-\frac{m\omega}{2\hbar}\left[x+b\cos\left(\omega t\right)\right]^{2}},\\
\\
S\left(x,t\right) & =m\omega b\left[x\sin\left(\omega t\right)+\frac{1}{4}\hbar a^{2}b\sin\left(2\omega t\right)\right]-\frac{1}{2}\hbar\omega t,
\end{cases}
\end{equation}
and the other from $a=-\sqrt{m\omega/\hbar}$:

\begin{equation}
\begin{cases}
R\left(x,t\right) & =\left(\frac{m\omega}{\pi\hbar}\right)^{\frac{1}{4}}e^{-\frac{m\omega}{2\hbar}\left[x-b\cos\left(\omega t\right)\right]^{2}},\\
\\
S\left(x,t\right) & -m\omega b\left[x\sin\left(\omega t\right)-\frac{1}{4}\hbar a^{2}b\sin\left(2\omega t\right)\right]-\frac{1}{2}\hbar\omega t,
\end{cases}
\end{equation}
which correspond, as expected \citep{Cohen}, to both coherent states

\begin{equation}
\psi_{\pm}\left(x,t\right)=\left(\frac{m\omega}{\pi\hbar}\right)^{\frac{1}{4}}e^{-\frac{m\omega}{2\hbar}\left\{ x^{2}+b\left[b\cos\left(\omega t\right)\mp2x\right]e^{-i\omega t}+i\frac{\hbar}{m}t\right\} }\label{aux18}
\end{equation}

Therefore, starting only from the general shape of the probability
density and assuming a behavior consistent with the features of the
harmonic potential, we were able to determine from the Bohmian equations
the required parameters $a$ and $b$, the first assuming only two
possible values and the latter any real value. Although we have considered
a Gaussian probability density (\ref{Pgauss}) that led to the coherent
state (\ref{aux18}), the method can be applied to more general densities
$P\left(x,t,\left\{ a_{i}\right\} \right)$, where $\left\{ a_{i}\right\} $
is a set of parameters to be determined considering the potential
$V\left(x,t\right)$ (in the previous example, we have $a_{1}\leftrightarrow a$
and $a_{2}\leftrightarrow b$), enabling the quantum state engineering
\citep{Orlowski,Paul,Vogel,Joshi}. However, it is not always possible
to associate a probability density with a given \emph{a priori} potential,
and we will illustrate this with the following example. Still considering
the harmonic potential, we will try to construct a Gaussian state
whose center remains stopped at the same point in space, with the
oscillatory character of the harmonic potential manifesting itself
in a periodic variation of its width: 
\begin{equation}
P\left(x,t\right)=\frac{1}{\sqrt{\pi}\left|b\right|}\frac{e^{-\left\{ \frac{x-a}{b\left[1+\varepsilon\sin\left(\omega t\right)\right]}\right\} ^{2}}}{1+\varepsilon\sin\left(\omega t\right)}\label{Pxt}
\end{equation}
Here, $a,\,b\,\varepsilon\in\mathbb{R}$ and $\left|\varepsilon\right|<1$
and all these parameters are left to be determined. Following the
method, we start with the continuity equation for $p\left(x,t\right)$,
Eq. (\ref{ContEq}), whose solution is given by

\begin{equation}
p\left(x,t\right)=m\omega\varepsilon\frac{\cos\left(\omega t\right)}{1+\varepsilon\sin\left(\omega t\right)}\left(x-a\right)+m\omega a\Theta\left(t\right)e^{\frac{x\left(x-2a\right)}{b^{2}\left[1+\varepsilon\sin\left(\omega t\right)\right]^{2}}}\label{pxt}
\end{equation}
where the boundary condition $\Theta\left(t\right)$ remains to be
determined. The Bohm's potential for Eq. (\ref{Pxt}) becomes 
\begin{align}
V_{B}\left(x,t\right) & =\frac{1}{2}m\omega^{2}x^{2}-\frac{\hbar^{2}}{2mb^{2}}\left\{ \frac{\left(x-a\right)^{2}}{b^{2}\left[1+\varepsilon\sin\left(\omega t\right)\right]^{4}}-\frac{1}{\left[1+\varepsilon\sin\left(\omega t\right)\right]^{2}}\right\} \label{VB}
\end{align}
and by substituting Eqs. (\ref{Pxt}), (\ref{pxt}) and (\ref{VB})
in the spatial derivative of the Hamilton-Jacobi equation (\ref{HJdiffs}),
we note that the only admissible condition is $\Theta\left(t\right)\equiv0$.
Consequently, from the canonical momentum definition, it follows the
action 
\begin{equation}
S\left(x,t\right)=\frac{m\omega\varepsilon}{2}\frac{\cos\left(\omega t\right)}{1+\varepsilon\sin\left(\omega t\right)}x\left(x-2a\right)+f\left(t\right)\label{Snovo}
\end{equation}
The analysis of the Hamilton-Jacobi equations (\ref{eqBhom1}) and
(\ref{HJdiffs}), we will find that it is not possible a valid solution
for both $b$ and $\varepsilon$ time-independent, therefore we conclude
that Eq. (\ref{Pxt}) is not a possible probability density to obtain
for the harmonic potential. However, it is known in the scientific
literature \citep{Pavicic,Pavicic1} the problem of investigating
which Hamiltonians possess Gaussians as their solutions and, then,
starting with (\ref{Pxt}) or other convenient Gaussian behavior and
maintaining $V\left(x,t\right)$ initially undefined, we can follow
the method to find an adequate $V\left(x,t\right)$ and define the
entire Hamiltonian - a possible application of the time-dependent
potential generated is on quantum computing by qubit arrays \citep{Almeida}.

\section{\textit{Conclusions} }

In this paper we presented a method, based on the Bohmian equations,
to reconstruct a wave-function $\psi\left(\vec{r},t\right)$ only
from the probability density $\left\vert \psi\left(\vec{r},t\right)\right\vert ^{2}$
and the potential energy of the system $V\left(\vec{r},t\right)$,
an approach that resembles some prior proposals \citep{Bardroff,Leonhardt}.
The determined final state $\psi\left(\vec{r},t\right)$ has no phase
ambiguities, except by a constant factor $e^{if\left(0\right)}$,
$f\left(0\right)\in\mathbb{R}$. The strength of our analytical method
is that it is entirely based on the Schrödinger equation (\ref{eqSchr3D})
in its totally equivalent form given by Bohm \citep{Bohm1} where
the magnitude and phase of the wave function are separated from one
another by construction. We have here an associated innovative character
since, to the continuity equation already used in previous approaches
\citep{Aerts,Paganin}, we now add the Hamilton-Jacobi equation. Since
the phase-retrieval problem dealt here is based different assumptions
from those widely discussed \citep{Orlowski,Corbett1,Weigert,Reichenbach,Moroz,Pauli,Gale,Richter,Corbett,Kemble},
our method circumvents the limitations pointed out by the cited authors.

In addition to practical applications in general \citep{Fienup,Gerchberg,Oppenheim,Zernike1,Zernike2,Gabor,Paganin}
and quantum physics \citep{Nakajima,Liberman,Vaccaro,Bardroff,Raymer},
we can raise issues from the fundamental point of view. Pauli emphasized
\citep{Pauli} that the wave functions themselves are not directly
observable and only constitute a mathematical tool to establish $P_{r}\left(\vec{r},t\right)$
and its relation with $P_{p}\left(\overrightarrow{p},t\right)=\left\vert \widetilde{\psi}\left(\overrightarrow{p},t\right)\right\vert ^{2}$.
However, there were recent attempts to measure the wave-function itself
\citep{Raymer,Lundeen} and, with the method presented here, we can
have a $\psi$-meter, as defined in Ref. \citep{Raymer}, if we measure
$\left\vert \psi\left(\vec{r},t\right)\right\vert ^{2}$. 

Our examples are all unidimensional, but the structure of the Bohm's
equations allows for generalization to two or three dimensions, as
suggested by Holland \citep{Holland}. Our next step is to test the
methodology we have developed with more challenging problems like
Non-Gaussian probability densities and multidimensional spaces.\textcolor{red}{{}
}The form of the Hamilton-Jacobi resulting equation depends on the
potential $V_{B}\left(\vec{r},t\right)$ form, and it can be difficult
to give a general analytic method to solve it \citep{FetterWalecka},
but there are advanced techniques \citep{Arnold,Vedenyapin,Romano}
that can be used when necessary, even through numerical implementations
\citep{Romano}. There are also different approaches to the continuity
equation \citep{FetterWalecka,Holland,Arnold,Romano,Clement}, mainly
in fluid dynamics \citep{Lamb,Landau}. Even the pure-states presented
here are not a limitation, because there are studies to expand the
Bohm's equation to mixed states \citep{Durr,Maroney}, and even to
open quantum systems \citep{Lorenzen,Nassar}. 

It is worth stressing that our method can be used for state engineering
\citep{Orlowski,Vogel,Joshi} under some known potential $V\left(\vec{r},t\right)$,
or for Hamiltonian engineering \citep{Pavicic,Vogel,Pavicic1,Almeida}
starting from a general form for the desired probability density $P_{r}\left(\vec{r},t,\left\{ a_{i}\right\} \right)$,
with applications on quantum computing \citep{Almeida} and quantum
optics \citep{Vogel}. The fact that our method is analytical in its
foundations and offers a clear answer to a fundamental question of
quantum theory, involving the determination of its most basic entity,
gives it a broad appeal. 
\begin{acknowledgments}
The authors wish to thank the Brazilian National Institute for Science
and Technology of Quantum Information (INCT-IQ) from Conselho Nacional
de Desenvolvimento Científico e Tecnológico (CNPq). C. A. Brasil wishes
to thank the Programa Nacional de Pós-Doutorado (PNPD) from Coordenação
de Aperfeiçoamento de Pessoal de Ensino Superior (CAPES). R. de J.
Napolitano wishes to thank the suport from Fundação de Amparo à Pesquisa
do Estado de São Paulo (FAPESP) project number 2018/00796-3 and also
from CNPq INCT-IQ 465469/2014-0. 
\end{acknowledgments}

\end{document}